\documentclass[useAMS,usenatbib]{mn2e}

\usepackage{graphicx}

\title[3C\,120: precession and variability]{Can long-term periodic variability and jet helicity in 3C\,120 be explained by jet precession?}
\author[A. Caproni and Z. Abraham]{A. Caproni$^{1}$\thanks{E-mail:
acaproni@astro.iag.usp.br} and Z. Abraham$^{1}$\\
$^{1}$Instituto de Astronomia, Geof\'\i sica e Ci\^encias Atmosf\'ericas, Universidade de S\~ao Paulo, PO Box 3386, 01060-970, S\~ao Paulo, Brazil}
\begin{document}

\date{}

\pagerange{\pageref{firstpage}--\pageref{lastpage}} \pubyear{2002}

\maketitle

\label{firstpage}

\begin{abstract}
   Optical variability of 3C\,120 is discussed in the framework of jet precession. Specifically, we 
   assume that the observed long-term periodic variability is produced by the emission from an 
   underlying jet with a time-dependent boosting factor driven by precession. The differences in 
   the apparent velocities of the different superluminal components in the milliarcsecond jet can 
   also be explained by the precession model as being related to changes in the viewing angle. The 
   evolution of the jet components has been used to determine the parameters of the precession model, 
   which also reproduce the helical structure seen at large scales. Among the possible mechanisms 
   that could produce jet precession, we consider that 3C\,120 harbours a super-massive black hole 
   binary system in its nuclear region and that torques induced by misalignment between the accretion 
   disc and the orbital plane of the secondary black hole are responsible for this precession; we 
   estimated upper and lower limits for the black holes masses and their mean separation.
\end{abstract}


\section{Introduction}

   3C\,120 (z=0.033; \citealt{bal80}), also known as II Zw 14 and PKS 0430+052, is usually 
   classified as a Seyfert 1 galaxy, although its morphology in the optical band is not as simple as  
   that of a typical galaxy of this class. Indeed, photometric and spectroscopic studies seem to indicate that 
   3C\,120 either passed, or it is still passing through a merger process (e.g., \citealt{sou89,hjo95}). 
   The residual I-band image obtained by \citet{hjo95} after subtraction of the stellar contribution,
   showed a complex structure formed by several condensations, probably associated to active star forming regions \citep{sou89}, 
   and an elongated structure which coincides with the kilo-parsec radio jet detected at 5 GHz \citep{wal88}.
   In fact, this large scale jet is the extension of a sub-parsec scale jet, which remains relativistic 
   up to distances of about 100 kpc. \citep{wal87a}. 
   
   Several superluminal radio components have been detected in the jet (e.g., \citealt{gom98,gom00,wal01,gom01}), 
   with different velocities and position angles, besides a stationary core smaller than 54 $\mu$as (0.025$h^{-1}$ pc) 
   \citep{gom99}. There is also evidence of the existence of trailing shocks in the jet \citep{gom01}, that is, 
   features that do not originate in the jet inlet. These features could be related to pinch-mode jet-body instabilities 
   produced by the propagation of the superluminal components, as shown recently by numerical simulations (e.g., 
   \citealt{agu01,alo03}). The observed correlation between dips in the X-ray emission and ejection of superluminal 
   components has been interpreted as a consequence of the connection between  jet origin and the accretion disc, 
   such as in the case of microquasars \citep{mar02}.

   3C\,120 presents variability in all bands and in different time-scales (e.g., \citealt{eps72,halp85,webb90,shst96,zdgr01}).
   The complex variability found in the historical {\it B}-band light curve was decomposed by \citet{webb90} 
   in three different components: a linear time decrease of its magnitude due to the diminution of the 
   accretion rate, a long-term variability with a period of 12.4-yr produced by thermal or viscous instabilities 
   in the accretion disc, and short-term variations associated to magnetic eruptions in the magnetized disc.
   
   In this paper, the reported 12.4-yr  variability is interpreted as periodic boosting of the radiation emitted by 
   the underlying jet, caused by jet precession. This model also explains the differences in superluminal velocities 
   and position angles of the different components, assuming that they represent the direction of the jet inlet at the 
   epoch in which the components were formed. Furthermore, the complex jet structure of 3C\,120 at large scales is 
   also studied in the framework of jet precession. Jet precession has been claimed by several 
   authors in order to explain the radio structure of several quasars, BL Lacs and radio-galaxies (e.g., 
   \citealt{gohu82,gow82,gohu84,rome87,abca98,abro99,abra00,sti03,caab03}), suggesting that jet precession is not so uncommon 
   phenomenon in the Universe. 
   We shall adopt $h=0.7$ and $q_{0}=0.5$ throughout the paper.

%\section{Superluminal jet knots and jet precession}

\section{Jet precession}

\subsection{Observational evidences}

   High-resolution observations between 5 and 43 GHz \citep{wal82,wal87b,gom99,gom00,fom00,hom01,wal01} give
   the core-component distance $r$ and the position angle on the plane of sky $\eta$ for each feature in 
   the sub-parsec scale jet. Using data at different epochs, the apparent proper motion $\mu$ is calculated and 
   the apparent velocity $\beta_{\mathrm{app}}$ in units of light speed $c$ is obtained  from: 

   \begin{eqnarray}
      \beta_{\mathrm{app}} = \frac{q_{0}z+(q_{0}-1)\lbrack(1+2q_{0}z)^{1/2}-1\rbrack}{100hq_{0}^2(1+z)}\mu
   \end{eqnarray}
   \\where $h=H_{0}/100$, $H_{0}$ is the Hubble constant in units of km s$^{-1}$ Mpc$^{-1}$, $q_{0}$ is the 
   deceleration parameter and $z$ is the redshift. The ejection epoch $t_{\mathrm{0}}$ of each component is 
   obtained by back-extrapolation of their linear motions. The kinematic parameters of the different superluminal 
   features are presented in Table 1.
   
   %-----------------------------------------------------------TABLE 01 

\begin{table*}
 \centering
 \begin{minipage}{140mm}
  \caption{Kinematic parameters of the superluminal components of 3C\,120.}
  \begin{tabular}{@{}cccccc@{}}
  \hline
   Component & Literature\footnote{Nomenclature in previous papers: $^{\mathrm{[1]}}$\citet{wal01}; $^{\mathrm{[2]}}$\citet{gom99}); 
                                $^{\mathrm{[3]}}$\citet{hom01}; $^{\mathrm{[4]}}$\citet{gom00}; $^{\mathrm{[5]}}$\citet{gom01}.}
 & $t_{\mathrm{0}}$ (yr) & $\mu$ (mas/yr) & $h\beta_{\mathrm{app}}$  & $\eta$ ($\degr$)\\
 \hline
K1                 & -                                            & 1976.7~$\pm$~0.6 & 3.01~$\pm$~0.33 & 4.6~$\pm$~0.5 & -113~$\pm$~2\\
K2                 & -                                            & 1977.6~$\pm$~0.6 & 3.01~$\pm$~0.39 & 4.6~$\pm$~0.6 & -108~$\pm$~4\\
K3                 & A$^{\mathrm{[1]}}$                           & 1978.8~$\pm$~0.5 & 2.75~$\pm$~0.33 & 4.2~$\pm$~0.5 & -101~$\pm$~9\\
K4                 & B$^{\mathrm{[1]}}$                           & 1980.3~$\pm$~0.5 & 3.01~$\pm$~0.26 & 4.6~$\pm$~0.4 & -99~$\pm$~5\\
K5                 & C$^{\mathrm{[1]}}$                           & 1981.0~$\pm$~0.5 & 2.48~$\pm$~0.26 & 3.8~$\pm$~0.4 & -95~$\pm$~6\\
K6                 & D$^{\mathrm{[1]}}$                           & 1981.7~$\pm$~0.4 & 2.35~$\pm$~0.20 & 3.6~$\pm$~0.3 & -97~$\pm$~6\\
K7                 & E$^{\mathrm{[1]}}$                           & 1982.6~$\pm$~0.4 & 2.22~$\pm$~0.20 & 3.4~$\pm$~0.3 & -101~$\pm$~7\\
K8                 & F$^{\mathrm{[1]}}$                           & 1983.3~$\pm$~0.5 & 2.09~$\pm$~0.20 & 3.2~$\pm$~0.3 & -102~$\pm$~8\\
K9                 & G$^{\mathrm{[1]}}$                           & 1984.3~$\pm$~0.5 & 2.09~$\pm$~0.39 & 3.2~$\pm$~0.6 & -109~$\pm$~8\\
K10                & H+I$^{\mathrm{[1]}}$                         & 1985.4~$\pm$~0.8 & 2.42~$\pm$~0.39 & 3.7~$\pm$~0.6 & -119~$\pm$~5\\
K11                & I$^{\mathrm{[1]}}$                           & 1986.1~$\pm$~0.6 & 2.81~$\pm$~0.33 & 4.3~$\pm$~0.5 & -120~$\pm$~6\\
K12                & J$^{\mathrm{[1]}}$                           & 1986.7~$\pm$~0.6 & 2.35~$\pm$~0.33 & 3.6~$\pm$~0.5 & -120~$\pm$~7\\
K13                & K$^{\mathrm{[1]}}$                           & 1988.0~$\pm$~0.4 & 2.81~$\pm$~0.39 & 4.3~$\pm$~0.6 & -117~$\pm$~7\\
K14                & A$^{\mathrm{[2]}}$                           & 1994.5~$\pm$~0.7 & 2.22~$\pm$~0.26 & 3.4~$\pm$~0.4 & -108~$\pm$~3\\
K15                & B$^{\mathrm{[2]}}$, K1A/U1A$^{\mathrm{[3]}}$ & 1994.9~$\pm$~0.7 & 2.16~$\pm$~0.26 & 3.3~$\pm$~0.4 & -106~$\pm$~3\\
K16                & C$^{\mathrm{[2]}}$, K1B/U1B$^{\mathrm{[3]}}$ & 1995.2~$\pm$~0.6 & 2.09~$\pm$~0.26 & 3.2~$\pm$~0.4 & -111~$\pm$~3\\
K17                & D$^{\mathrm{[2,4]}}$, d$^{\mathrm{[2,4]}}$   & 1995.5~$\pm$~0.4 & 2.16~$\pm$~0.20 & 3.3~$\pm$~0.3 & -113~$\pm$~3\\
K18                & G2$^{\mathrm{[2]}}$, g$^{\mathrm{[2]}}$      & 1996.3~$\pm$~0.5 & 2.09~$\pm$~0.26 & 3.2~$\pm$~0.4 & -119~$\pm$~5\\
K19                & H$^{\mathrm{[2,4]}}$, h$^{\mathrm{[2,5]}}$   & 1996.8~$\pm$~0.4 & 2.09~$\pm$~0.20 & 3.2~$\pm$~0.3 & -120~$\pm$~5\\
K20                & J$^{\mathrm{[2,4]}}$, j$^{\mathrm{[2]}}$     & 1997.0~$\pm$~0.4 & 2.16~$\pm$~0.26 & 3.3~$\pm$~0.4 & -116~$\pm$~3\\
K21                & K$^{\mathrm{[2,4]}}$, k$^{\mathrm{[5]}}$     & 1997.3~$\pm$~0.4 & 2.22~$\pm$~0.20 & 3.4~$\pm$~0.3 & -117~$\pm$~5\\
K22                & L$^{\mathrm{[2,4]}}$, l2$^{\mathrm{[5]}}$    & 1997.5~$\pm$~0.4 & 2.29~$\pm$~0.20 & 3.5~$\pm$~0.3 & -124~$\pm$~4\\
K23                & o1+o2$^{\mathrm{[5]}}$                       & 1998.2~$\pm$~0.4 & 2.22~$\pm$~0.13 & 3.4~$\pm$~0.2 & -122~$\pm$~3\\
\hline
\end{tabular}
\end{minipage}
\end{table*}

%______________________________________________________________

   We have labelled jet components as `K' followed by a number related to the epoch in which they were formed 
   (`1' for the oldest one). We also present the labels given in earlier works in the second column of Table 1. 
   Except for K10, we kept strictly the previous identifications.  Considering the uncertainties, the listed 
   parameters are compatible with those found in the literature (e.g., \citealt{gom98,gom01,wal01}).

   We can note in Table 1 that the different jet components were ejected with different velocities and position angles. A 
   possible interpretation for this behaviour is in terms of a precessing jet model: precession changes the orientation 
   of the jet inlet in relation to the line of sight, so that the direction in which the components are ejected, 
   as well as their apparent velocities become a function of time. For the present discussion, it is not relevant 
   whether the jet components are assumed to be plasmons or shocks since we are only interested in the kinematic 
   aspects. 
   
    At lower resolution,  VLBI observations at 1.7 GHz have revealed that the jet structure of 3C\,120 is extremely 
   complex \citep{wal01}; it presents sub-structures in scales of tenths of a parsec that move superluminally, a 
   possible stationary component located at an angular distance of about 81 mas from the core and a jet aperture 
   that is larger in the southern direction, specially after about 180 mas, where  there is also a decrease in the 
   apparent velocity.
   
   Those characteristics were interpreted by \citet{wal01} as an indicative of the presence of a helical pattern 
   in the jet. However, they can also be understood in terms of the precession model that explains the sub-parsec 
   behavior of the superluminal jet.
   
   \subsection{Precession model}
 
   We will derive the instantaneous appearance of the jet, assuming that it is the result of the 
   combination of plasma elements ejected in different epochs, with different angles in relation to the 
   line of sight. Let us consider a plasma element ejected at time $t_0$ with velocity $c\beta$ 
   ($c$ is the light speed) in the comoving reference frame. This element will have a velocity 
   $c\beta_{\mathrm{app}}$ in the observer's reference frame given by:

   \begin{eqnarray}
      \beta_{\mathrm{app}} = \frac{\pm\beta\sin\lbrack\phi(t_0)\rbrack}{1\pm\beta\cos\lbrack\phi(t_0)\rbrack} 
   \end{eqnarray}
   \\where $\phi$ is the angle between the moving direction of the element and the line of sight. The signs 
   `+' and `-' refer to the jet and the counterjet respectively. As the counterjet has not been detected for 
   scales smaller than 100 kpc \citep{wal87a}, we will consider only the jet hereafter. 

   Due to jet precession, $\phi$ and $\eta$ are functions of time $t$ given by:

   \begin{eqnarray}
%      \phi(t_0) = \arcsin\lbrack\sqrt{x(t_0)^2+y(t_0)^2}\rbrack 
      \phi(t) = \arcsin\biggl[{\sqrt{x(t)^2+y(t)^2}}\biggr] 
   \end{eqnarray}
   
      \begin{eqnarray}
      \eta(t) = \arctan\biggl[\frac{y(t)}{x(t)}\biggr]
   \end{eqnarray}
   \\with 
   
   \begin{eqnarray}
     x(t) = A(t)\cos\eta_0-B(t)\sin\eta_0
   \end{eqnarray}

   \begin{eqnarray}
     y(t) = A(t)\sin\eta_0+B(t)\cos\eta_0
   \end{eqnarray}   
   \\and

   \begin{eqnarray}
     A(t) = \cos\Omega\sin\phi_0+\sin\Omega\cos\phi_0\sin\omega t
   \end{eqnarray}

   \begin{eqnarray}
     B(t) = \sin\Omega\cos\omega t
   \end{eqnarray}
   \\where $\omega$ is the precession angular velocity, $\Omega$ is the semi-aperture angle of the precession 
   cone, $\phi_{0}$ is the angle between the precession cone axis and the line of sight and $\eta_{0}$ the 
   projected angle of the cone axis on the plane of the sky.
   
   In the determination of the precession parameters, we used all jet components listed in Table 1 as model 
   constraints. As data were obtained at different frequencies, shifts in the core-component distances due 
   to opacity effects may become quantitatively important in the calculation of the proper motions. To perform 
   opacity correction in the observational data, we used the formalism given in \citet{blko79}, which uses the 
   integrated synchrotron luminosity $L_{\mathrm{syn}}$, the ratio between upper and lower limits of the energy 
   distribution in the relativistic jet particles $\Gamma_{\mathrm{max}}/\Gamma_{\mathrm{min}}$, the intrinsic 
   jet aperture angle $\psi^\prime$, a constant parameter $k_{\mathrm{e}}$ and the angle $\phi$ between the jet 
   direction and the line of sight. The relation between these quantities is presented in Appendix A.
   
   As in \citet{loba98}, we assumed $k_{\mathrm{e}}=1$, $\Gamma_{\mathrm{max}}/\Gamma_{\mathrm{min}}=100$, 
   $\psi^\prime=0.5\degr$ and $L_{\mathrm{syn}}=8.4\times10^{41}$ erg s$^{-1}$. However, in our model, the angle 
   $\phi$ is a function of time  and depends on the precession model parameters. Therefore, the corrections were
   determined iteratively together with the model. The final result was a small correction, with
   a mean value  between 5 and 43 GHz of $\Delta r_{\mathrm{core}}=0.147$ mas, with  upper and lower limits 
   of 0.130 and 0.164 mas, respectively.
    
   To find the precession parameters we  consider a  period of 12.3 yr\footnote{In the 
   source's reference frame, the precession period corresponds to 11.9 yr}, almost the same as the long-term 
   variability period found by \citet{webb90} in the {\it B}-band light curve. In order to determine the best set of model 
   parameters, we chose a $\gamma$ factor compatible with the velocity of the fastest component ($\sim4.6 h^{-1}$ c), 
   given by $\gamma_{min}=(1+\beta_{\mathrm{app}}^2)^{1/2}$ ($\gamma_{min}\sim6.7$ for $h=0.7$). After fixing 
   a value for $\gamma$ close to its lower limit, we selected the parameters $\Omega$, $\phi_{\mathrm{0}}$ and 
   $\eta_{\mathrm{0}}$ that fitted the apparent velocities and position angles of the jet components. Then, we 
   checked the behavior of $h\beta_{\mathrm{app}}$ and $\eta$ as functions of time, assuming that the position 
   angle of the jet component represents the jet position at the epoch when the component was formed. This procedure 
   was done iteratively until a good fitting for the data was obtained.

   The precession parameters are given in Table 2, while the model fitting in the ($\eta$, $h\beta_{\mathrm{app}}$), 
   ($t$, $h\beta_{\mathrm{app}}$) and ($t$, $\eta$) planes is presented in Fig. 1. 
   It is also possible to fit the data using larger values for $\gamma$ and other jet parameters 
   (decreasing $\Omega$ and $\phi_{\mathrm{0}}$). As a consequence of that, not only 
   the predicted position angles are much larger than those observed in the VLBI maps but also the
    time variations of the Doppler boosting factor become smaller than those necessary to explain the optical
   light curve (see Section 4 for further discussion). 

%-----------------------------------------------------------TABLE 02 

\begin{table}
  \caption{Parameters of the precession model for the parsec-jet of 3C\,120.}
  \begin{tabular}{@{}ccccc@{}}
  \hline
  $P$ (yr)$^{\mathrm{a}}$ & $\gamma$ & $\Omega$ ($\degr$) & $\phi_{\mathrm{0}}$ ($\degr$) & $\eta_{\mathrm{0}}$ ($\degr$)\\
 \hline
12.3~$\pm$~0.3 & 6.8~$\pm$~0.5 & 1.5~$\pm$~0.3 & 4.8~$\pm$~0.5 & -108~$\pm$~4\\
\hline
\end{tabular}
      \begin{list}{}{}
         \item[$^{\mathrm{a}}$] Measured in the framework fixed at the observer. 
      \end{list}
\end{table}

%______________________________________________________________

%-----------------------------------------------------------FIGURE 01 
   \begin{figure}
      {\includegraphics{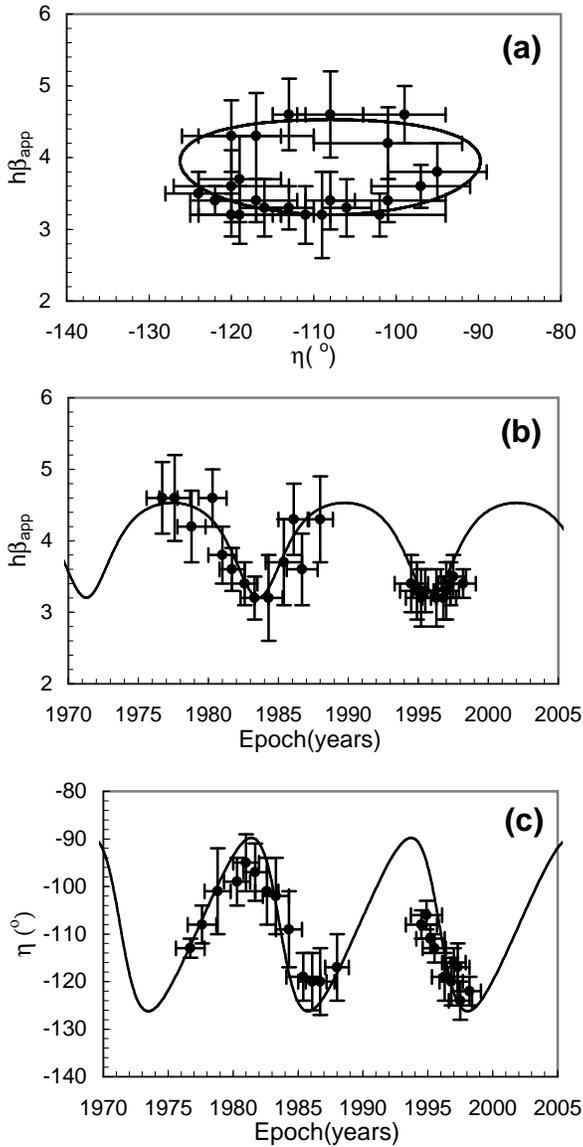}}
      \caption{Precession model for 3C\,120. The parameters are listed in Table 2. {\bf (a) - (c):} Solutions of 
               the model on the planes ($\eta$, $\mathrm{h}\beta_{\mathrm{app}}$), ($t$, $\mathrm{h}\beta_{\mathrm{app}}$) 
               and ($t$, $\eta$). 
             }
         \label{precession model}
   \end{figure}

%______________________________________________________________

\section{Large scale jet structure of 3C 120}

  Taking into account that even though the jet components have ballistic trajectories on the plane of the sky, 
  a snapshot in time of a precessing jet will show an helicoidal pattern. To compare our results with the 
  large-scale jet structure described by \citet{wal01}, we calculated the apparent proper motion $\mu$ of 
  a jet element from $c\beta_{\mathrm{app}}[\phi(t)]$ through equation (1).

  Using equations (2)-(8), we determined the right ascension and declination offsets of the jet element in 
  relation to the core ($\Delta\alpha$ and $\Delta\delta$ respectively) in a given time $t_{\mathrm{obs}}$ 
  ($t_{\mathrm{obs}}\geq t_0$) through:

   \begin{eqnarray}
     \Delta\alpha(t_{\mathrm{obs}}) = \mu_{\alpha}(t_0)\cdot (t_{\mathrm{obs}}-t_0)
   \end{eqnarray}

   \begin{eqnarray}
     \Delta\delta(t_{\mathrm{obs}}) = \mu_{\delta}(t_0)\cdot (t_{\mathrm{obs}}-t_0)
   \end{eqnarray}
   \\where $\mu_{\alpha}$ and $\mu_{\delta}$ are respectively the apparent proper motions in right ascension 
   and declination, being related to $\mu$ and $\eta$ by:

   \begin{eqnarray}
     \mu_{\alpha}(t_0) = \mu(t_0)\cdot \sin[\eta(t_0)]
   \end{eqnarray}

   \begin{eqnarray}
     \mu_{\delta}(t_0) = \mu(t_0)\cdot \cos[\eta(t_0)]
   \end{eqnarray}

   Assuming now a continuous jet and the precession parameters listed in Table 2, we simulated the jet appearance at 
   the five distinct epochs for which 1.7 GHz observations are available \citep{wal01}: 
   $t_{\mathrm{obs}}$ = 1982.77, 1984.26, 1989.85, 
   1994.44 and 1997.70, as shown in Fig. 2.
   Note that this approach is similar to that used to model the radio structure of 
   several quasars and radio-galaxies in previous papers \citep{gohu82,gow82,gohu84,rome87}.
   
  %-----------------------------------------------------------FIGURE 02 
   \begin{figure}
      {\includegraphics{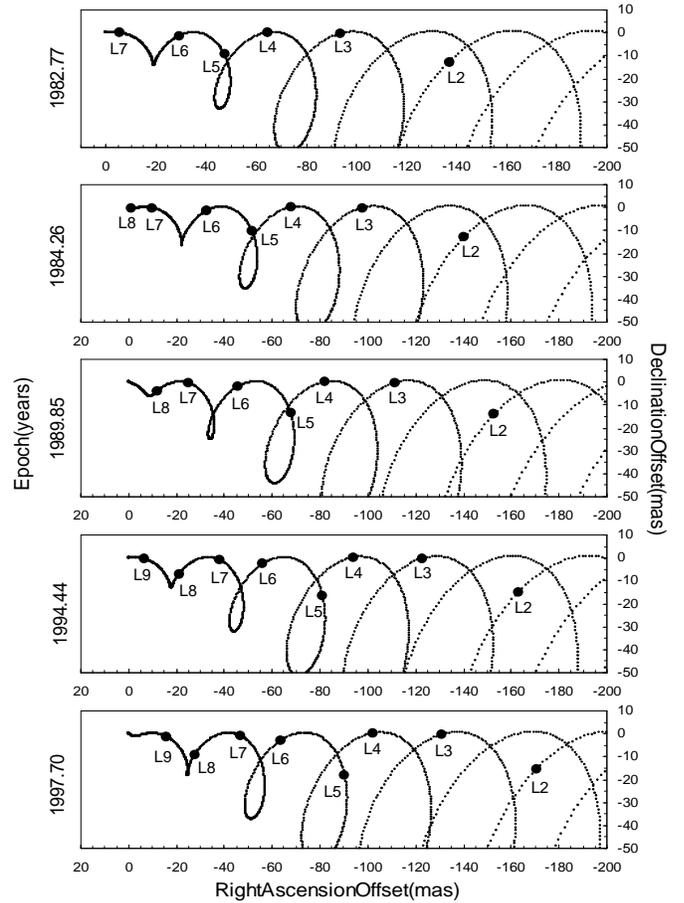}}
      \caption{Helicoidal pattern due to jet precession. Each one of the panels corresponds to snapshots of the precession helix 
               in five distinct epochs, chosen as the same epochs of the 1.7 GHz observations \citep{wal01}. The big circles 
               superposed upon the helices represent the position of jet components which can be related to the features L2-L9 
               found by \citet{wal01}. 
              }
         \label{jet helicity}
   \end{figure}

%______________________________________________________________

   Each point of the helices in Fig. 2 corresponds to a given plasma element, although in the case of a continuous jet, 
   the helix driven by precession is  continuous. The wavelength of the helices is basically related to the precession 
   period, while the amplitude depends mainly on $\Omega$ and $\phi_{\mathrm{0}}$. As this pattern is not stationary in 
   time, since the precession helix is tied up with the jet movement, the configuration observed in a given time $t$ 
   will be seen again after an interval $\Delta t$, which corresponds to 12.3 yr in the case of 3C\,120.

    The comparison of the predicted helicoidal patterns in Fig. 2 with the VLBI maps obtained by \citet{wal01} show 
   that the observed jet aperture is reasonably well described by jet precession in the inner parts of the jet.
	  The  maps reveal also the existence of discrete structures labelled as L2-L9, which 
   could be superluminal. In order to compare these observations with our model, we assumed that L2-L9 
   were ejected at different epochs and have ballistic motions. We present in Table 3 their kinematic parameters, 
   which are used to calculated their right ascension and declination offsets for the five epochs given in Fig. 2 
   (their positions are marked by big circles). Components L7 and L8 can be identified respectively as the evolved 
   components K5 and K8 in the parsec-scale jet, as presented in Table 1.

%-----------------------------------------------------------TABLE 04 

\begin{table}
  \caption{Kinematic parameters of the superluminal components observed at 1.7 GHz in the large scale jet of 3C\,120.}
  \begin{tabular}{@{}ccccc@{}}
  \hline
  Component & $t_{\mathrm{0}}$ (yr) & $\mu$ (mas/yr) & $h\beta_{\mathrm{app}}$  & $\eta$ ($\degr$)\\
 \hline
L2 & 1921.3 & 2.24 & 3.4 & -95\\
L3 & 1945.0 & 2.49 & 3.8 & -90\\
L4 & 1957.2 & 2.52 & 3.9 & -90\\
L5 & 1966.4 & 2.94 & 4.5 & -101\\
L6 & 1970.2 & 2.32 & 3.5 & -93\\
L7 & 1980.7 & 2.76 & 4.2 & -91\\
L8 & 1983.7 & 2.10 & 3.2 & -108\\
L9 & 1992.1 & 2.88 & 4.4 & -95\\
\hline
\end{tabular}
\end{table}

%______________________________________________________________

   A good agreement between the offsets shown in Fig. 2 and the locations of L2-L9 in the 1.7 GHz maps of 3C\,120 is 
   found, suggesting that the simple approach assumed in this work (ballistic motion + precession) provide a reasonable 
   description for their kinematic behaviour in such scales. However, this conclusion could be somehow misleading, since 
   we can rule out neither the possibility that L2-L9 are formed by superposition of several unresolved components 
   in the maps, nor that jet components have non-ballistic trajectories in those scales.  
  
 At larger distances from the core ($\mid\Delta\alpha\mid>80$ mas), we see that 
   the full range of position angles provided by our precession model is not found in the observational data, 
   in the sense that the amplitudes of the precession helices towards the southern direction are 
   systematically larger than the observed jet aperture. A possible explanation for this behavior is the existence of an 
   external medium which does not allow the jet propagation in some directions and could lead to the formation
   of a stationary component, as observed in the VLBI maps
   at $\Delta\alpha\approx -80$ mas and $\Delta\delta\approx -17$ mas.
   
   From these results, it is reasonable to verify in what conditions the  quasi-stationary component could be formed.
   The interaction between the fluids at different velocities will produce shock waves propagating along the jet.
   In the case of strong shocks and assuming that the fluid can be described by a relativistic adiabatic equation 
   of state ($p\propto N^{4/3}$, where $p$ is the thermodynamic pressure), 
   the jump condition between the pre-shock and 
   pos-shock regions is (e.g., \citealt{blmc76,rom96}):

   \begin{eqnarray}
      \frac{N_{\mathrm{ps}}}{N_{\mathrm{j}}} = 4\gamma_{\mathrm{ps}}\gamma \biggl[1-\sqrt{1-\gamma_{\mathrm{ps}}^{-2}-\gamma^{-2}+(\gamma_{\mathrm{ps}}\gamma)^{-2}}\biggr]+3
   \end{eqnarray}
   \\where $N_{\mathrm{j}}$ and  $\gamma$ are respectively the proper particle density and the bulk Lorentz 
   factor in the pre-shock region, while $N_{\mathrm{ps}}$  and $\gamma_{\mathrm{ps}}$ are the same quantities in  the  
   pos-shock region.

   If the observed feature 
   is actually stationary, its proper motion must not produce displacements larger than the angular resolution of 
   the maps, otherwise its motion would have been detected during this interval. From the beam-size in right ascension 
   and declination (4 and 12.5 mas respectively) and the interval between the last and first observation ($\sim$14.93 yr), 
   we estimated an upper limit for the apparent proper motion of about 0.27 mas/yr, which results in an apparent 
   velocity smaller than 0.41$c$. 

   To calculate the true velocity of the quasi-stationary knot, it is necessary to know its viewing angle. However, 
   only its position angle (about $-102\degr$) is known and the precession model allows two different solutions 
   for the viewing angle: 3.3$\degr$ and 6.2$\degr$. Using equation (2) with $\phi=6.2\degr$, we found that the upper 
   limits for the true velocity and the associated Lorentz factor for the quasi-stationary component are respectively 
   $0.80c$ and $1.67$, while for $\phi=3.3\degr$ the same quantities have upper limits of $0.88c$ and $2.11$.

   With these limits and assuming that the quasi-stationary component is associated with the pos-shock region, 
   we found from  equation (13) that the minimum particle density required to slow down the jet 
   is $12N_{\mathrm{j}}$ and $10N_{\mathrm{j}}$ for the viewing angles $6.2\degr$ and 
   $3.3\degr$, respectively.  For an external medium like the 
   clouds of the Narrow Line Region, with  typical number density of 10$^4$ cm$^{-3}$, and for 
   a viewing angle of 6.2$\degr$, we obtain an upper limit of
    of 833 cm$^{-3}$ for the jet density,  which is reasonable in 
   terms of ordinary relativistic jets (e.g., \citealt{wal87a,alt89,rom96}).

\section{The optical boosted emission of the underlying jet}

   The luminosity of 3C\,120 observed in the optical band might be the result of the superposition of  thermal 
   contribution from the accretion disc with non-thermal contribution produced, e.g., in the underlying jet and in 
   the superluminal components. If we consider that the underlying jet is represented by a relativistic continuous 
   fluid, its flux density measured in the observer's reference frame $S_{\mathrm{j}}(\nu)$ is related to that in 
   the comoving frame $S_{\mathrm{j}}^{\prime}(\nu)$ by \citep{libl85}:

   \begin{eqnarray}
      S_{\mathrm{j}}(\nu) = S_{\mathrm{j}}^{\prime}(\nu)\delta(\phi,\gamma)^{2+\alpha}
   \end{eqnarray}
   \\where $\alpha$ is the spectral index ($S_{\nu}\propto\nu^{-\alpha}$) and $\delta$ is the Doppler 
   factor, defined as: 

   \begin{eqnarray}
      \delta(\phi,\gamma) = \lbrack\gamma(1-\beta\cos\phi)\rbrack^{-1} 
   \end{eqnarray}

   Due to  jet precession, $\delta$,   and consequently the observed emission from the underlying jet, become time 
   dependent; the closer the jet is to the line of sight, the higher the boosting effect is. 

   Depending on the intrinsic jet intensity, periodic boosting variations could produce detectable variability in 
   the light curve, as in the case of 3C\,279 and OJ\,287 \citep{abca98,abra00}. To reproduce this light curve and 
   following the suggestion by \citet{webb90} of a possible secular linear variation, we modeled the contribution 
   of the underlying jet assuming that its emission is described by equation (14), with $S_{\mathrm{j}}^{\prime}(\nu)$ 
   given by:

   \begin{eqnarray}
      S_{\mathrm{j}}^{\prime}(\nu,t) = S_{\mathrm{1}}^{\prime}(\nu)+S_{\mathrm{2}}^{\prime}(\nu)\cdot (t-1965)\cdot \frac{\delta(\phi,\gamma)}{(1+z)}
   \end{eqnarray}

   The term $\delta(\phi,\gamma)/(1+z)$ in equation (16) is due to the transformation of the time measured in the comoving 
   frame to the observer's reference frame. Combining equations (14) and (16), we have:

   \begin{eqnarray}
      S_{\mathrm{j}}(\nu,t) = \biggl[S_{\mathrm{1}}^{\prime}(\nu)+S_{\mathrm{2}}^{\prime}(\nu)\cdot (t-1965)\cdot\nonumber\\\cdot \frac{\delta(\phi,\gamma)}{(1+z)}\biggr]\cdot \delta(\phi,\gamma)^{2+\alpha}
   \end{eqnarray}

   Observations in the {\it B} and {\it I} bands of the optical counterpart of the kilo-parsec radio jet led to 
   $\alpha_{B-I}=2.0$ \citep{hjo95}; assuming that this value  is also valid for the parsec-scale 
   jet and using  $\delta$ from the precession model, we  simulated the long-term periodic variability 
   in two different cases: Model A, with a secular time decrease of the intrinsic flux density, and 
   Model B, in which the linear term was neglected ($S_{\mathrm{2}}^{\prime}=0$). The model parameters 
   are given in Table 4, where  $S_{\mathrm{1}}^{\prime}(\nu)$ was chosen to 
   keep $S_{\mathrm{j}}(\nu,t)$ smaller than the minimum observed value at all epochs. For that reason 
   the values of the parameters  should be considered as upper limits. 

%-----------------------------------------------------------TABLE 03 

\begin{table}
  \caption{Model parameters used in equation (17) to describe long-term variability in the {\it B}-band light curve of 3C\,120.}
  \begin{tabular}{@{}ccccc@{}}
  \hline
  & & $S_{\mathrm{1}}^{\prime}$ ($\mu$Jy) & & $S_{\mathrm{2}}^{\prime}$ ($n$Jy)\\
 \hline
MODEL A & & 0.19 & & -0.14 \\
MODEL B & & 0.16 & & 0.0 \\
\hline
\end{tabular}
\end{table}

%______________________________________________________________

%-----------------------------------------------------------FIGURE 03 
   \begin{figure}
      {\includegraphics{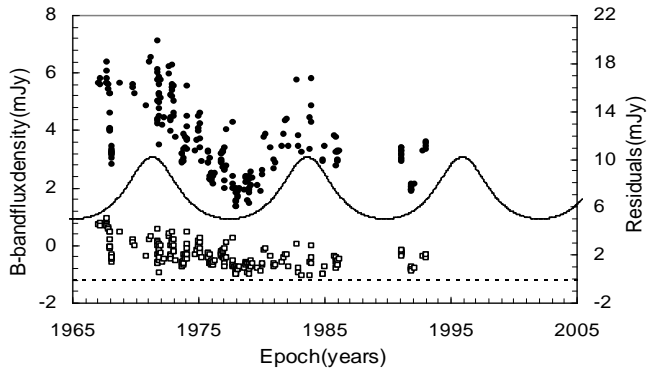}}
      \caption{Contribution of the underlying jet for the variability in the {\it B}-band light curve of 3C\,120 in the 
               case of Model B (see Table 4). The full circles are related to the observed flux density, while open 
               squares are the residuals obtained after subtracting the contribution of the underlying jet (solid lines). 
               The dashed line corresponds to residual equals to zero.
              }
         \label{correlations between components and flares}
   \end{figure}

%______________________________________________________________

   Boosting effects produce a substantial increase in the underlying jet flux density in both models. Since 
   the residuals obtained from the subtraction of the observed flux density from the model predictions are 
   very similar in the two cases, the secular linear decrease in the {\it B}-band brightness proposed originally 
   by \citet{webb90} seems to be unnecessary, at least after 1965. We show in Fig. 3 the flux density calculated 
   from model B, the {\it B}-band historical light curve of 3C\,120 and the difference between them.

\section{A possible super-massive black hole binary system in the nuclear region of 3C\,120}

   Jet precession can be produced by the Lense-Thirring effect 
   \citep{leth18}, in which precession is due to the misalignment between the angular momenta of the accretion 
   disk and of a Kerr black hole. This effect was investigated by \citet{lu92} for several AGNs; for all 
   of them, the period found was of the order of thousands of years. Using the same assumptions for
   the central object in 3C\,120, which has a central mass of $3.4\times10^7$ M$_{\sun}$ \citep{pet98}, we obtain
   an even longer precession period.
   On the other hand, jet precession with periods of several years can be produced in super-massive black hole binary 
   systems, when 
   the secondary black hole has an orbit non-coplanar with the primary accretion disc, which induces torques in 
   its inner parts (e.g., \citealt{katz80,katz97,rom00})
   Thus, we will assume that the former scenario, in which jet inlet precession is induced in super-massive black hole 
   binary system, is more suitable to 3C\,120 and, from this assumption, the binary system parameters will be estimated. 

   Let us consider that the primary and secondary black holes, with masses $M_{\mathrm{p}}$ and $M_{\mathrm{s}}$ 
   respectively, are separated by a distance $r_{\mathrm{ps}}$. From Kepler's third law, we can relate $r_{\mathrm{ps}}$ 
   to the orbital period of the secondary around the primary black hole $P_{\mathrm{ps}}$ through:

   \begin{eqnarray}
      r_{\mathrm{ps}}^3 = \frac{\mathrm{G}M_{\mathrm{tot}}}{4\pi^2}P_{\mathrm{ps}}^2
   \end{eqnarray}
   \\where G is the gravitational constant and $M_{\mathrm{tot}}$ is the sum of the masses of the two 
   black holes. 
   
   In the observer's reference frame, the orbital period $P_{\mathrm{ps}}^{obs}$ is given by:
   
   \begin{eqnarray}
      P_{\mathrm{ps}} = \frac{P_{\mathrm{ps}}^{obs}}{(1+z)}
   \end{eqnarray}

   According to \citet{rom00}, density waves, which disturb the accretion rate and originate the superluminal components, 
   could be produced when the secondary black hole crosses the primary accretion disc. As the secondary crosses the disc 
   twice per orbit, we used $P_{\mathrm{ps}}^{obs}$ as twice the mean separation between the emergence of superluminal 
   components, which leads to $P_{\mathrm{ps}}\approx 1.4$ yr.

   Using reverberation mapping techniques, \citet{pet98} estimated a virial mass for the central source of 
   $3.4\times 10^7$ M$_{\sun}$. We assumed that this value corresponds to the total mass of the binary system 
   and from equation (18), with the values of $P_{\mathrm{ps}}$ and $M_{\mathrm{tot}}$ given above, we found 
   that $r_{\mathrm{ps}}\approx 5.9\times10^{15}$ cm. Considering that the outer radius of the precessing part 
   of the disc is $r_{\mathrm{d}}$, \citet{pate95} and \citet{larw97} calculated its precession period $P_{\mathrm{d}}$ 
   in terms of the masses of the black holes:

   \begin{eqnarray}
     \frac{2\pi}{P_{\mathrm{d}}}(1+z) = -\frac{3}{4}\left(\frac{7-2n}{5-n}\right) \frac{GM_{\mathrm{s}}}{r_{\mathrm{ps}}^3}\times\nonumber\\\times\frac{r_{\mathrm{d}}^2}{\sqrt{GM_{\mathrm{p}}r_{\mathrm{d}}}}\cos\theta
   \end{eqnarray}
   \\where $n$ is the politropic index of the gas (e.g., $n=3/2$  and $n=3$ for the non-relativistic
   and relativistic cases, respectively) and $\theta$ is the angle between  the orbit of the secondary and 
   the plane of the disc. 

   If the jet and accretion disc are coupled, the jet  precesses at same rate than the disc ($P=P_{\mathrm{d}})$, 
   forming a precession cone with half-opening angle equal to the angle of orbit inclination ($\Omega=\theta$). 
   Thus, replacing $M_{\mathrm{s}}$ by $M_{\mathrm{tot}}-M_{\mathrm{p}}$ in equation (20), we can calculate 
   $r_{\mathrm{d}}$ in terms of $M_{\mathrm{p}}$ and $M_{\mathrm{tot}}$:

   \begin{eqnarray}
      r_{\mathrm{d}} = \left[-\frac{8\pi}{3}\left(\frac{5-n}{7-2n}\right)\frac{(1+z)}{P\cos\Omega}\frac{r_{\mathrm{ps}}^3}{\sqrt{GM_{\mathrm{tot}}}}\right]^{2/3}\times\nonumber\\\times\frac{x_{\mathrm{p}}^{1/3}}{\left(1-x_{\mathrm{p}}\right)^{2/3}}
   \end{eqnarray}
   \\where $x_{\mathrm{p}}=M_{\mathrm{p}}/M_{\mathrm{tot}}$. In Fig. 4, we plot $r_{\mathrm{d}}$ as a function of 
   $x_{\mathrm{p}}$ using the precession parameters listed in Table 2. We can observe that the increase of 
   $r_{\mathrm{d}}$ with $x_{\mathrm{p}}$ is more pronounced after $x_{\mathrm{p}}\approx 0.7$, such that little variations in $x_{\mathrm{p}}$ 
   introduces large changes in $r_{\mathrm{d}}$.

%-----------------------------------------------------------FIGURE 04 
   \begin{figure}
      {\includegraphics{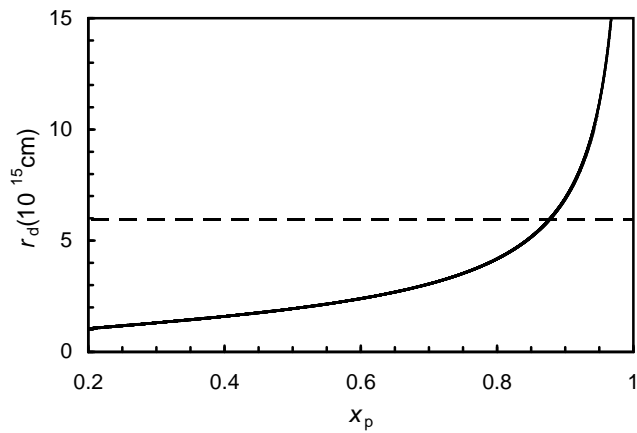}}
      \caption{Outer radius of the precessing disc as a function of the fractional mass of the primary black hole (full line). The dashed line 
                 indicates the separation between the primary and secondary black holes.   
              }
         \label{Outer radius of the precessing disc}
   \end{figure}

%______________________________________________________________

  This formalism is valid only if the disc precesses as a rigid-body, implying that $r_{\mathrm{d}}$ 
  must be appreciably smaller than $r_{\mathrm{ps}}$ \citep{pate95}. This limit provides an additional 
  constrain to the masses of the black holes. The dashed line in Fig. 4 shows the value of $r_{\mathrm{ps}}$ 
  derived from equation (18). In order to satisfy the hypothesis of rigid-body precession, the allowed solutions 
  are found for values of $r_{\mathrm{d}}$ which lie under the dashed line in Fig. 4. It necessarily means 
  that $x_{\mathrm{p}}<0.88$, so that $M_{\mathrm{p}}<3.0\times 10^7 M_{\sun}$. As 
  $M_{\mathrm{tot}}=M_{\mathrm{p}}+M_{\mathrm{s}}$, we can also obtain an lower limit for the mass of the 
  secondary, which should be higher than $4.0\times 10^6 M_{\sun}$. Summarizing, we present in Table 5 the 
  parameters of the black hole binary system calculated in this section.

%-----------------------------------------------------------TABLE 05 

\begin{table}
  \caption{Parameters of a possible black hole binary system in the inner parts of 3C\,120.}
  \begin{tabular}{@{}ccccc@{}}
  \hline
  $P_{\mathrm{ps}}$ (yr) & $r_{\mathrm{ps}}$ (cm) & $M_{\mathrm{p}}$ (M$_{\sun}$)$^{\mathrm{a}}$ & $M_{\mathrm{s}}$ (M$_{\sun}$)$^{\mathrm{b}}$ & $M_{\mathrm{tot}}$ (M$_{\sun}$)$^{\mathrm{c}}$\\
 \hline
1.4 & $5.9\times10^{15}$ & $3.0\times10^7$ & $4.0\times10^6$ & $3.4\times10^7$\\
\hline
\end{tabular}
      \begin{list}{}{}
         \item[$^{\mathrm{a}}$] Upper limit; 
         \item[$^{\mathrm{b}}$] Lower limit; 
         \item[$^{\mathrm{c}}$] \citet{pet98}. 
      \end{list}
\end{table}

%______________________________________________________________

\section{Conclusions}

   Based on the periodicity in the historical {\it B}-band light curve and variable jet structure, 
   we propose the existence of jet precession in 3C\,120, with a period of 12.3 yr. 

   We assume that the different apparent velocities of the superluminal components measured at milliarcsecond 
   and larger scales are related to changes in the angle between the jet inlet and the line of sight due to 
   precession, although the superposition of unresolved components and/or interaction with the environment 
   could are acting at the largest scales. 

   We show that the periodicity in the optical light curve can be produced by the boosted underlying jet, 
   with a time-dependent boosting factor driven by precession. An upper limit of 0.16 $\mu$Jy was estimated 
   for the flux density of the underlying jet in the comoving reference frame. The inclusion of a secular linear 
   term in the analysis of the long-term variability, as in \citet{webb90}, is not necessary to obtain a good 
   fitting to the light curve.

   The helicoidal jet pattern  found by \citet{wal01} is interpreted in this work also as the result of jet 
   precession. The helix generated by precession reproduces quite well the jet aperture seen in the 1.7-GHz maps 
   up to distances from the core smaller than $\sim$80 mas, where there is a probable stationary component. 
   Beyond that, the helix amplitude is systematically larger in the southern direction, suggesting the existence 
   of an external medium that does not allow jet propagation. In order to produce a stationary component, 
   considering a one-dimensional adiabatic relativistic jet as well as energy and particle flux conservation,
   we estimate a lower limit of $\sim$12 for the ratio between jet and environment densities. 
   
   Assuming that jet precession in 3C\,120 is driven by a secondary super-massive black hole in a non-coplanar 
   orbit around the primary accretion disc, using the total mass of the two black holes derived from reverberation 
   mapping techniques and an orbital period of approximately 1.4 yr, we estimate an upper limit of $3.0\times10^7$ M$_{\sun}$ 
   for the primary black hole mass, a lower limit of $4.0\times10^6$ M$_{\sun}$ for the secondary mass and a 
   separation between them of about $5.9\times10^{15}$ cm.

\section*{Acknowledgments}

      This work was supported by the Brazilian Agencies FAPESP (Proc. 99/10343-3), CNPq and FINEP. We would like 
      to thank the anonymous referee for her/his useful comments and suggestions.

\appendix

\section[]{Opacity effects on core-component distance and precession jet}

   As it was pointed out previously (e.g., \citealt{blko79,loba96,loba98}), the absolute core position 
   $r_{\mathrm{core}}$ depends inversely on the frequency when the core is optically thick, what introduces 
   a shift in the core-component separation. Following \citet{blko79}, we can write the absolute core 
   position as:  

   \begin{eqnarray}
      r_{\mathrm{core}}(\nu) = \frac{4.56\times 10^{-12}(1+z)}{D_{\mathrm{L}}\gamma^2 k_{\mathrm{e}}^{1/3}\psi \sin\phi}\cdot\nonumber\\\cdot \left[\frac{L_{\mathrm{syn}}\sin\phi}{\beta(1-\beta\cos\phi)\ln(\Gamma_{\mathrm{max}}/\Gamma_{\mathrm{min}})}\right]^{2/3}\nu^{-1} (\mathrm{mas})
   \end{eqnarray}
   \\where $z$ is the redshift, $D_{\mathrm{L}}$ is the luminosity distance (in units of parsec), 
   $k_{\mathrm{e}}$ is a constant ($k_{\mathrm{e}}\leq 1$; \citealt{blko79}), $L_{\mathrm{syn}}$ 
   is the integrated synchrotron luminosity (in units of erg s$^{-1}$), while $\Gamma_{\mathrm{max}}$ 
   and $\Gamma_{\mathrm{min}}$ are related respectively to the upper and lower limits of the energy 
   distribution of the relativistic jet particles. The quantities $\psi$ and $\nu$ are respectively 
   the observed aperture angle of the jet (in radians) and the frequency (in Hz); the former is 
   related to the intrinsic jet aperture angle $\psi^\prime$ through (e.g., \citealt{mut90}):

   \begin{eqnarray}
      \tan(\psi/2) = \tan(\psi^\prime/2)\cot\phi 
   \end{eqnarray}  

   The core position shift $\Delta r_{\mathrm{core}}$ between frequencies $\nu_1$ and $\nu_2$ 
   ($\nu_2\geq\nu_1$) is given by:

   \begin{eqnarray}
      \Delta r_{\mathrm{core}}(\nu_1,\nu_2) = \frac{4.56\times 10^{-12}(1+z)}{D_{\mathrm{L}}\gamma^2 k_{\mathrm{e}}^{1/3}\psi \sin\phi}\cdot\nonumber\\\cdot \left[\frac{L_{\mathrm{syn}}\sin\phi}{\beta(1-\beta\cos\phi)\ln(\Gamma_{\mathrm{max}}/\Gamma_{\mathrm{min}})}\right]^{2/3}\frac{(\nu_2-\nu_1)}{\nu_1\nu_2} (\mathrm{mas})
   \end{eqnarray}

   Note that if we substitute in equation (A3) a simpler version of equation (A2), $\psi\approx \psi^\prime\csc\theta$, 
   we obtain equation (11) given in \citet{loba98}.

   We can see that equation (A3) depends on the angle between the jet and line of sight; in the case 
   of a jet which is precessing, this angle is a function of time, what obviously introduces a 
   time dependency in $\Delta r_{\mathrm{core}}$. On the other hand, the shifts 
   in the core-component separations do not occur in a fixed direction, but they are oriented 
   according to the direction whose jet inlet is pointed. As the jet inlet is not resolved by 
   observations, changes in its direction will reflect on changes in the position angle of the 
   core region. Thus, a jet component, located at a distance $r$ from the core and with a position 
   angle $\eta$, will have right ascension and declination offsets ($\Delta\alpha$ and $\Delta\delta$ 
   respectively) given by:

   \begin{eqnarray}
     \Delta\alpha(\nu_1,\nu_2) = r(\nu_1)\sin[\eta(t_{\mathrm{obs}})]-\nonumber\\-\Delta r_{\mathrm{core}}(\nu_1,\nu_2)\sin[\eta_{\mathrm{c}}(t_{\mathrm{obs}})]
   \end{eqnarray}

   \begin{eqnarray}
     \Delta\delta(\nu_1,\nu_2) = r(\nu_1)\cos[\eta(t_{\mathrm{obs}})]-\nonumber\\-\Delta r_{\mathrm{core}}(\nu_1,\nu_2)\cos[\eta_{\mathrm{c}}(t_{\mathrm{obs}})]
   \end{eqnarray}
   \\where $\eta_{\mathrm{c}}$ is the position angle of the core in the epoch $t_{\mathrm{obs}}$ in 
   which observation is acquired. Known the precession model parameters, we are able to determine the 
   second term of the equations (A4a) and (A4b) and correct the component position by core opacity effects; 
   using them, we can determine the corrected core-component distance $r_{\mathrm{corr}}$ 
   through:

   \begin{eqnarray}
      r_{\mathrm{corr}}(\nu_1) = [r(\nu_1)^2+\Delta r_{\mathrm{core}}(\nu_1,\nu_2)^2-\nonumber\\-2r(\nu_1)\Delta r_{\mathrm{core}}(\nu_1,\nu_2)\cos(\eta-\eta_{\mathrm{c}})]^{1/2}
   \end{eqnarray}

   Note that if $\Delta r_{\mathrm{core}}=0$, $r_{\mathrm{corr}}=r$. Other particular case 
   is found when there is alignment between the position angles of the core and of the jet 
   component ($\eta=\eta_{\mathrm{c}}$), such that 
   $r_{\mathrm{corr}}=r-\Delta r_{\mathrm{core}}$.

\section[]{Glossary}

   In order to facilitate the reading of the manuscript, we define all the symbols 
   that appear in the text in Table B1.

%-----------------------------------------------------------TABLE B1 

\begin{table*}
 \centering
 \begin{minipage}{140mm}
  \caption{Definition of the symbols used in the text.}
  \begin{tabular}{@{}ll@{}}
  \hline
  Symbol & Meaning\\
 \hline
$G$                        & gravitational constant \\
$c$                        & light speed \\
$H_{0}$                    & Hubble constant \\
$h$                        & $H_{0}/100$ \\
$q_{0}$                    & deceleration parameter \\
$z$                        & redshift \\
$D_{\mathrm{L}}$           & luminosity distance \\
$r$                        & core-component distance \\
$\Delta\alpha$             & right ascension offset of components relative to the core position \\
$\Delta\delta$             & declination offset of components relative to the core position\\
$t_{\mathrm{0}}$           & ejection epoch for the jet components\\
$\eta$                     & position angle on the plane of the sky of jet components \\
$\mu$                      & apparent proper motion of jet components\\
$\beta_{\mathrm{app}}$     & apparent velocity of jet components \\
$\phi$                     & viewing angle of jet components\\
$\omega$                   & jet precession angular velocity \\
$P$                        & jet precession period \\
$\Omega$                   & semi-aperture angle of the precession cone \\
$\phi_{\mathrm{0}}$        & angle between the precession cone axis and the line of sight \\
$\eta_{\mathrm{0}}$        & projected angle of the precession cone axis on the plane of the sky \\
$\beta$                    & jet bulk velocity \\
$\gamma$                   & Lorentz factor of the jet bulk motion \\
$\delta$                   & Doppler factor of jet components\\
$\gamma_{\mathrm{min}}$    & lower limit for the jet bulk motion \\
$\gamma_{\mathrm{ps}}$     & Bulk Lorentz factor of the pos-shock region \\
$p$                        & thermodynamic pressure \\
$N_{\mathrm{j}}$           & proper particle density of the jet \\
$N_{\mathrm{ps}}$          & proper particle density of the pos-shock region \\
$n$                        & politropic index of the gas \\
$\nu$                      & frequency \\
$S_{\mathrm{j}}$           & flux density of the underlying jet in the observer's reference frame \\
$S_{\mathrm{j}}^\prime$    & flux density of the underlying jet in the comoving reference frame \\
$\alpha$                   & flux density spectral index \\
$M_{\mathrm{p}}$           & mass of the primary black hole \\
$M_{\mathrm{s}}$           & mass of the secondary black hole \\
$M_{\mathrm{tot}}$         & total mass inside  the nuclear region \\
$x_{\mathrm{p}}$           & ratio between the primary  and the total masses \\
$r_{\mathrm{ps}}$          & separation between the primary and secondary black holes \\
$P_{\mathrm{ps}}^{obs}$    & orbital period of the secondary around the primary black hole in the observer's reference frame\\
$P_{\mathrm{ps}}$          & orbital period of the secondary around the primary black hole in the source's reference frame\\ 
$r_{\mathrm{d}}$           & outer radius of the precessing part of the primary accretion disc \\
$P_{\mathrm{d}}$           & precession period of the accretion disk \\
$\theta$                   & angle between the orbital plane of the secondary and the plane of the primary disc \\
$L_{\mathrm{syn}}$         & integrated synchrotron luminosity \\
$\Gamma_{\mathrm{max}}$    & upper limit for the energy distribution of relativistic jet particles \\
$\Gamma_{\mathrm{min}}$    & lower limit for the energy distribution of relativistic jet particles \\
$\psi^\prime$              & intrinsic jet aperture angle \\
$\psi$                     & observed jet aperture angle \\
$k_{\mathrm{e}}$           & constant parameter \\
$\Delta r_{\mathrm{core}}$ & angular shift in core position \\
$\eta_{\mathrm{c}}$        & core position angle \\
\hline
\end{tabular}
\end{minipage}
\end{table*}

%______________________________________________________________

\bsp

\label{lastpage}

\end{document}